\documentclass[prd,preprintnumbers,superscriptaddress,floatfix,twocolumn,titenlines,notitlepage,nofootinbib,showpacs]{revtex4-1}
\pdfoutput=1
\usepackage{textcase}
\usepackage{graphicx,epsfig,psfrag,amssymb}
\usepackage{multirow}
\usepackage{color,xcolor,graphicx,epsfig,psfrag,amsmath,empheq}
\usepackage{bm}
\usepackage{mathrsfs, amsfonts, color}
\usepackage{hepunits}
\usepackage[colorlinks=true,citecolor=blue,linkcolor=blue, urlcolor=.]{hyperref}
\usepackage{cleveref}
\usepackage{slashed}
\usepackage[caption=false]{subfig}
\usepackage{xspace}
\usepackage{hepunits}
\usepackage{multirow}
\usepackage{enumitem}
\usepackage{verbatim}
\usepackage[utf8]{inputenc}
\setlist[description]{leftmargin=\parindent,labelindent=\parindent}

\definecolor{darkerblue}{rgb}{0.2,0.2,0.5}
\definecolor{seagreen}{rgb}{0.180392,0.545098,0.341176}

\usepackage{tikz}
\usetikzlibrary{trees}
\usetikzlibrary{decorations.pathmorphing}
\usetikzlibrary{decorations.markings}
\tikzset{
    photon/.style={decorate, line width=0.15mm, decoration={snake,amplitude=3pt,segment length=8pt}, draw=black},
    wino/.style={draw=redwine},    
    fermion/.style={draw=black, line width=0.2mm, postaction={decorate},
        decoration={markings,mark=at position .55 with {\arrow[draw=black,scale=2,#1]{>}}}},
    scalar/.style={draw=black, dashed,postaction={decorate},
        decoration={markings,mark=at position .55 with {\arrow[draw=black,scale=2,#1]{>}}}},
    scalarline/.style={draw=black, postaction={decorate},
        decoration={markings,mark=at position .55 with {\arrow[draw=black,scale=2,#1]{>}}}},
    gluon/.style={decorate, draw=black,
        decoration={coil,amplitude=3pt, segment length=4pt}},
    graviton/.style={decorate, draw=black,
        decoration={zigzag,amplitude=3pt, segment length=4pt}}
}
\tikzstyle{blob}=[circle,
                              minimum size=20,
                              draw=black!80,
                              fill=black!80]   
\tikzstyle{redblob}=[circle,
                              thick,
                              minimum size=0.4cm,
                              draw=red!80,
                              fill=red!60]

\newcommand{\be}{\begin{eqnarray}}
\newcommand{\ee}{\end{eqnarray}}
\def\lsim{\mathrel{\rlap{\lower4pt\hbox{\hskip 0.5 pt$\sim$}}
\raise1pt\hbox{$<$}}}

\newcommand{\tev}{\ensuremath{\mathrm{\: Te\kern -0.1em V}}\xspace}
\newcommand{\gev}{\ensuremath{\mathrm{\: Ge\kern -0.1em V}}\xspace}
\newcommand{\mev}{\ensuremath{\mathrm{\: Me\kern -0.1em V}}\xspace}

\def\Dbar    {{\kern 0.2em\overline{\kern -0.2em \mathrm{D}}{}}\xspace}

\def\kstarbar    {{\kern 0.2em\overline{\kern -0.2em K}{}^{*0}}\xspace}

\definecolor{darkgreen}{rgb}{0,0.5,0}
\definecolor{darkblue}{rgb}{0,0,0.5}
\definecolor{newred}{rgb}{0.5,0.1,0}
\definecolor{gold}{rgb}{0.7,0.7,0}
\definecolor{newpurple}{rgb}{0.5,0,0.5}

\begin{document}

\title{Co-Interacting Dark Matter  }

\author{Jia Liu}
\email{liuj1@uchicago.edu}
\affiliation{\mbox{Physics Department and Enrico Fermi Institute, University of Chicago, Chicago, IL 60637}}

\author{Xiao-Ping Wang}
\email{Corresponding author. xia.wang@anl.gov}
\affiliation{\mbox{High Energy Physics Division, Argonne National Laboratory, Argonne, IL 60439}}

\author{Wei Xue}
\email{wei.xue@cern.ch}
\affiliation{\mbox{Theoretical Physics Department, CERN, CH-1211 Geneva 23, Switzerland}}

\begin{abstract}
We propose a novel mechanism with two component dark matter models. The subdominant dark matter can thermalize the dominant one in galaxies, and leads to core density profiles. Using ultralight dark photons and $\mathrm{GeV}$-$\mathrm{TeV}$ Dirac fermions as an example, we couple the two dark matter candidates with a $U(1)$ interaction. 	
This mechanism differs from self-interacting dark matter, due to three effects: (1) higher occupation numbers, (2) forward-backward scattering cancellation, and (3) the multiple scatterings required for the heavy dark matter.
Furthermore, the Bullet Cluster bound is evaded due to the reduced Bose enhancement factor.
Unlike the fuzzy dark matter solution to the small structure problems which have tension with Lyman-$\alpha$, the ultralight dark photons with mass $\gtrsim 10^{-21}$ eV can have a core profile through interactions with $\psi$ and are not constrained by other astrophysical observations.
\end{abstract}

\preprint{EFI-19-3}
\preprint{CERN-TH-2019-010}
\maketitle

\section{Introduction} \label{sec:intro}
There is plenty of cosmological and astrophysical evidence of dark matter~(DM) and its gravitational 
interactions, yet its nature remains a mystery. Theoretically speaking, the mass of DM remains 
largely unknown. One well-motivated scenario is weakly interacting massive particles~(WIMP).
Another possibility is ultralight oscillating {\it fields} with a large occupation number in the universe. 
Candidates for this include the QCD axion \cite{PhysRevD.16.1791, Peccei:1977hh, PhysRevLett.40.223, Wilczek:1977pj, Preskill:1982cy, Abbott:1982af, Dine:1982ah, Kim:2008hd,Wantz:2009it, Ringwald:2012hr, Kawasaki:2013ae}, 
axion-like particles~(ALPs) \cite{Svrcek:2006yi,Essig:2013lka,  Marsh:2015xka, Graham:2015ouw}
and dark photons~\cite{Nelson:2011sf, Arias:2012az, Graham:2015rva}, etc.

Astrophysical observations provide clues on the properties of DM. This can shed light on the future research into dark matter. It has been established that the cold dark matter model can explain the large scale of the universe. 
However, tension in the small scale remain. This will either require a better understanding of the baryonic physics~\cite{Navarro:1996bv, Gelato:1998hb, Gnedin:2001ec, 0004-637X-560-2-636, 0004-637X-580-2-627, Ahn:2004xt,Tonini:2006gwz, Martizzi:2011aa, Read:2018fxs},
or imply new DM features. The small scale issues include the cusp/core problem
\cite{Flores:1994gz,Navarro:1996gj,0004-637X-765-1-25}, the missing satellite problem \cite{Klypin:1999uc},
and the too-big-to-fail problem
\cite{BoylanKolchin:2011de,BoylanKolchin:2011dk,Sand:2003bp,Tollerud:2014zha}. 
Explanations for these small scale observations through DM include self-interacting 
dark matter~(SIDM) \cite{Spergel:1999mh,Tulin:2017ara} with an interaction cross-section 
$\sigma / m \sim {\rm cm}^2/ {\rm g}$, and 
ultralight bosonic (fuzzy) DM~\cite{Hu:2000ke, Hui:2016ltb}
with the mass of order $10^{-22}~\mathrm{eV}$.
Baryonic effects can potentially solve these issues, but this remains unsettled. 
A recent study of isolated dwarf galaxies has shown that if there is  a core
in the halo, baryonic feedback will not lower the density profile further \cite{Fitts:2018ycl}.

The \textit{first motivation} of this work is to emphasize that \textit{any interactions} can push the system 
into equilibrium, thus redistributing the DM density in galaxies. This should not be limited to DM
self-interactions. One natural possibility is that a subdominant component of DM thermalizes 
the dominant one through interactions between them. In SIDM, two DM particles exchange 
their momentum in one collision. While in our scenario, two particles of the dominant DM separately 
scatter with the subdominant DM and exchange their momentum indirectly. The effect is the equivalent of 
the direct exchange. Thus, the DM density and velocity distribution
will reach equilibrium in the end, and the time cost will depend on the strength of the interactions and 
the density of  the system. In the center of galaxies, where the DM density is higher, 
the thermalization process is expected to be faster.

The scenario considered contains two components of DM. The self-interactions
are negligible, but the interactions between them are sizable enough to explain the small structure observations. 
We dub this scenario {\it Co-Interating Dark Matter}~(CoIDM).
One DM candidate is ultralight bosonic fields. 
The $10^{-22}~\mathrm{eV}$ bosonic fields as fuzzy DM can potentially solve the small structure issues,
due to their Broglie wavelength of $\sim \mathrm{kpc}$. 
However, the fuzzy DM is inconsistent with the Lyman-$\alpha$ constraints~\cite{Menci:2017nsr, Irsic:2017yje, Armengaud:2017nkf, Kobayashi:2017jcf, Murgia:2018now, Nori:2018pka}. 
For larger masses, the Lyman-$\alpha$ constraint is evaded, 
but its behavior is similar to the cold DM in galactic scales. Thus, the small scale problems remain. 
We will  show that by adding interactions with the other 
DM, \textit{ultralight fields as the dominant DM can have a core profile for masses 
larger than} $\sim 10^{-21}$ eV, thus solving the small 
structure issues. Meanwhile, the Bullet Cluster bound is avoided easily due to the reduced 
Bose enhancement factor. These points serve as the \textit{second motivation} for this work. 
\newline

\section{Models} \label{sec:idea}

We introduce a simple model for CoIDM. 
It contains an ultralight vector DM, a dark photon $A'$. Its relic abundance is achieved by 
non-thermal processes in the early universe, for instance, through inflationary quantum fluctuations~\cite{Graham:2015rva},
parametric resonances~\cite{Co:2018lka, Dror:2018pdh, Bastero-Gil:2018uel, Agrawal:2018vin}, and 
cosmic strings~\cite{Long:2019lwl}.
The other component is fermionic particles $\psi$, interacting with $A'$ via $U(1)$ gauge coupling $g'$,  
\begin{equation}
   \mathcal{L} \supset g' \bar{\psi} \gamma_{\mu } \psi A'^{\mu}   \  .
\end{equation}
The sum of the fractions of relic abundance for the two DM is the total DM abundance, 
$F_{A'} + F_\psi =1$. We have assumed the $\psi$ comes from interactions with
Standard Model particles via freeze-in, not from the $A'$ which is itself non-thermal. 
\newline

\section{Interaction rates and small scale structure}
The SIDM reaches its kinetic equilibrium in the center of galaxies, where it forms core density 
profiles~\cite{Balberg:2002ue,Ahn:2004xt,Kaplinghat:2015aga};
outside the central regions, it has less than one collision per particle in the galactic time scale, 
such that the density profile is similar to the collisionless DM. 
For CoIDM, after reaching the kinetic equilibrium, the equilibrium equations for the dominant DM 
are the same as the one for SIDM, such that 
%
CoIDM can have core profiles like SIDM. 
We analyze the interactions between $A'$ and $\psi$ to understand the dynamical time scale for the
dominant DM approaching equilibrium distributions. Two situations are considered: $A'$  and $\psi$ dominant.

The evolution of the phase space density functions $\mathcal{N}_{\psi,A'}$ is determined by the Boltzmann 
equation: 
\begin{equation}
   \left( \partial_t + v_i  \partial_{x_i}   + \dot{v}_i \partial_{v_i}  \right)  
         \mathcal{N} ( \bf{x}, \bf{p},  t ) = 
         \mathcal{C} ( \bf{x}, \bf{p},  t )   \ , 
   \label{eq:boltz}
\end{equation}
where $\mathcal{C}$ is the collision kernel deciding the time scale for $\mathcal{N}$ to reach
equilibrium.  The $\dot{v}_i$ term is proportional to the forces on $A'$ or $\psi$, which can come from the 
gravitational potentials or from the fields themselves. 
For the scattering process of $ \psi ( k_1 ) + A' ( p_1 ) \rightarrow \psi ( k_2 ) + A' (p_2 ) $, 
the leading collisional kernel for DM $\psi$ is 
\begin{eqnarray} 
   & \mathcal{C}_\psi  \simeq  
      \sum_{spin} \int \frac{ d^3 {\bf{p_1} }  d^3 {\bf{k_2}} } { ( 2 \pi )^5 8 m_{A'}^2 m_\psi^2  } 
            | \mathcal{M} \left( {\bf k_1}, {\bf p_1}, {\bf k_2},{\bf  p_2}  \right) |^2  \label{eq:collionK}  
         \label{eq:Boltz}
\\
         & 
        \times  \delta (  E_{k_1} + E_{p_1} - E_{k_2} - E_{p_2} )
         \mathcal{N}^{A'}_{p_1} \mathcal{N}^{A'}_{p_2} \left( \mathcal{N}^\psi_{k_2} - \mathcal{N}^\psi_{k_1} \right) 
         \nonumber  , 
\end{eqnarray} 
where the limits of the Bose enhancement $\mathcal{N}^{A'} \gg 1$ and the
non-relativistic DM are taken. The leading collisional kernel for dark photon $A'$ is the same as eq.~(\ref{eq:Boltz}), 
up to substituting $\int d^3 \bf{p_1}$ to $\int d^3 \bf{k_1}$,
\begin{eqnarray} 
& \mathcal{C}_{A'}  \simeq  
\sum_{spin} \int \frac{ d^3 {\bf{k_1} }  d^3 {\bf{k_2}} } { ( 2 \pi )^5 8 m_{A'}^2 m_\psi^2  } 
| \mathcal{M} \left( {\bf k_1}, {\bf p_1}, {\bf k_2},{\bf  p_2}  \right) |^2  
\label{eq:BoltzAp}
\\
& 
\times  \delta (  E_{k_1} + E_{p_1} - E_{k_2} - E_{p_2} )
\mathcal{N}^{A'}_{p_1} \mathcal{N}^{A'}_{p_2} \left( \mathcal{N}^\psi_{k_2} - \mathcal{N}^\psi_{k_1} \right) 
\nonumber  . 
\end{eqnarray} 

Due to the mass hierarchy $m_{A'} \ll m_\psi$ and the large occupancy number $\mathcal{N}^{A'} \gg 1$, the 
interaction rates have several features that distinguish them from ordinary particle scatterings:
\begin{itemize}
\item
{\it enhancement from the large occupation number in the final state.}
In the galaxies, the velocity dispersion $v_0$ of $A'$ is $\mathcal{O}(10^{-3})$, and 
$A'$ has a typical momentum of approximately $ m_{A'} v_0 $. Therefore, the occupation number 
for $A'$ is estimated as:
\begin{align} 
   & \langle \mathcal{N}^{A'} \rangle \sim \frac{ \rho_{A'} /m_{A'} } { m_{A'}^3 v_0^3 } 
         \sim 3\times 10^{76} \nonumber \\
   &  \times \left( \frac{\rho_{A'}} { 0.1 \mathrm{M_\odot/ pc^3}} \right)
         \left( \frac{ m_{A'}} {10^{-18} \mathrm{eV} } \right)^{-4}
         \left( \frac{ v_0} {10^{-3}  } \right)^{-3} .
         \label{eq:Na}
\end{align} 
The large occupation number is determined by the very small DM mass and galaxy formation, which fix the 
typical velocity $v_0 \sim 10^{-3} c$ and the local DM density. In this sense, it is similar as non-relativistic gas 
particles with fixed energy density.  When decreasing its mass, the number density increases accordingly.
We should mention that having a large occupation number does not necessarily mean a Bose-Einstein condensate (BEC). 
The ultralight dark photons in the non-relativistic limit are similar to scalars, such as axion dark matter. There have 
been plenty of studies on the time evolution of axion states in the galaxies, for example \cite{Hui:2016ltb}. 
After relaxation, the axion can form coherent oscillating soliton in the center of the galaxy, which is indeed BEC. 
But it only occupies a small fraction of the density, while most of them are still in the form of the axion gas with 
a high occupation number. Therefore, the dark photons are expected to share the same property.

The interaction with fermions cannot dramatically change the occupation number of $A'$, because it is 
the non-relativistic collision, its rate is small and the initial velocity distributions for $A'$ and $\psi$ are similar.
In more detail, after collisions with fermions, each dark photon can have order one change in the momentum. 
This is enough to transfer kinetic energy (heat) but does not change the fact that the majority of dark photon 
have velocity of the order of $v_0$. In this sense, the kinetic thermalized states are still non-relativistic. 

In terms of classical physics, the above process is the Thompson scattering with stimulated 
$A'$ emission. It is the dark photon version of the laser emission, in which the off-shell 
$\psi$ in the scattering diagram plays the role of excited atoms in laser physics.

\item
{\it suppression from the forward-backward scattering cancellation.}
In eq.~(\ref{eq:collionK}) and (\ref{eq:BoltzAp}), the collision kernel contains the cancellation 
from inverse scattering, specifically in the $\mathcal{N}^\psi_{k_2} - \mathcal{N}^\psi_{k_1}$ term. 
The momentum 
$\bf{k_1}$ and $\bf{k_2}$ are proximate because $m_\psi \gg m_{A'}$. 
The typical momentum of $\psi$ is $\sim m_\psi v_0$,  while the exchange $\Delta k$ is $ \sim m_{A'} v_0$ per collision due to momentum conservation. Since $\frac{d \mathcal{N}}{dk} \sim \mathcal{N}^\psi/(m_\psi v_0)$ for a smooth $\mathcal{N}^\psi$, we have 

\begin{align}\left( \mathcal{N}^\psi_{k_2} - \mathcal{N}^\psi_{k_1} \right) \sim \frac{d \mathcal{N}}{dk} \Delta k   \sim  \mathcal{N}^\psi \times \frac{m_{A'} } {m_\psi},
\end{align}

which contains one suppression factor $m_{A'} / m_\psi $. The above approximation requires $\mathcal{N}^\psi_{k_2} - \mathcal{N}^\psi_{k_1}$ to be non-zero, and it changes sign when switching $k_1$ and $k_2$.

\item
{\it suppression from multiple scattering requirements for $\psi$.}
Scattering once changes the momentum of $\psi$ by a small amount, $\sim m_{A'} v_0$. 
In order to thermalize $\psi$ and form a core density profile, the momentum change has to be $\sim \mathcal{O}(1) m_\psi v_0$. Therefore, multiple scatterings for $\psi$ are necessary,
and the number of ${A'}$-$\psi$ collisions should be around $\sim m_\psi^2 /m_{A'}^2$ as in the result
of random walking.
Therefore, the \textit{effective} interaction rates for $\psi$ should pay the penalty factor $m_{A'}^2 /m_\psi^2$ accordingly.
This argument 
does not apply for $A'$ (or SIDM), since one collision is normally enough to change the momentum of lighter 
(or equal mass) DM by $\mathcal{O}(1)$ factor. The other way to understand this multiple scattering 
suppression is to consider the momentum exchange rate, instead of the single ${A'}$-$\psi$ collision rate.
In this way, the suppression factor is automatically included.
\end{itemize}

Having $\psi$ and ${A'}$ with similar velocities in the galaxies, and
considering the suppression and enhancement effects above, 
the \textit{effective} interaction rate of $\psi$ is estimated as:  
\begin{align}
   \Gamma_\psi^{\rm eff} \simeq \frac{ m_{A'}^2}{m_\psi^2} \mathcal{C}_\psi
     \simeq n_{A'} \left\langle \sigma v \right\rangle_{\psi A' }    \langle \mathcal{N}^{A'} \rangle \frac{m_{A'}^3}{m_\psi^3},
      \label{eq:chi-rate}
\end{align}
where the cross-section for $\psi A' \to \psi A'$ scattering is:
\begin{equation} 
\left\langle \sigma v \right\rangle_{\psi A' }
\simeq \frac{g'^4 v_{\rm rel}}{4 \pi m_\psi^2}  \,,
\label{eq:Apsics}
\end{equation} 
and $ v_{\rm rel} $ is the relative velocity between $\psi$ and $A'$. 
The \textit{effective} interaction rate for ${A'}$ 
does not need multiple scatterings, 
\begin{equation}
   \Gamma_{A'}^{\rm eff}  \simeq n_\psi \left\langle 
   \sigma v \right\rangle_{\psi {A'}  }   \langle \mathcal{N}^{A'} \rangle  
   \left( \frac{m_{A'}}{m_\psi} \right) .
   \label{eq:a-rate}
\end{equation}
Due to the high ratio of $m_\psi / m_{A'}$, the collision rate for ${A'}$ is normally much larger than the collision 
rate for $\psi$, $\Gamma_{A'} \gg \Gamma_\psi$. 

To reach the kinetic equilibrium and form core profiles, the momentum exchange rate 
from $A'$-$\psi$ collision must satisfy 
$\Gamma \sim 0.1 \mathrm{Gyr}^{-1}$ for the dominant DM~\cite{Tulin:2017ara}. 
This depends on the density and velocity of DM in the galaxies.
In the central region of typical Dwarf galaxies, $\rho_{\rm DM} \sim 0.1 {\rm M}_{\odot}/{\rm pc}^3$,
and velocity dispersion $v_0 \sim 10 {\rm km}/{\rm s}$ \cite{ Walker:2009zp, Oh:2010ea}. 
When $A'$ and $\psi$ are in the same DM halo, 
$ v_{\rm rel} \sim v_0 $ is a good approximation.
We will use these requirements to map out the parameter spaces for the CoIDM model.
\\

1) \textit{${A'}$ dominant}, $ F_{A'} \approx 1 \gg F_\psi$. 
In Dwarf galaxies, the \textit{effective} collision rate for the dominant $A'$ is:
\begin{align}
& \Gamma_{A'}^{\rm eff} \approx  0.14 {\rm Gyr}^{-1}  \frac{F_\psi}{0.05} \left( \frac{g'}{10^{-12}} \right)^4
\left( \frac{m_{A'}}{10^{-18}\rm eV}\right)^{-3}  \left( \frac{m_{\psi}}{1 \rm GeV}\right)^{-4} \nonumber \\
& \times \left( \frac{v_{\rm rel} }{10 \rm{km/s} } \right) \left( \frac{v_0}{10 \rm{km/s}} \right)^{-3}
\left(\frac{\rho_{\rm DM}}{0.1{\rm M}_{\odot}/ {\rm pc}^3}\right)^2 .
\label{eq:GammaApnum }
\end{align}
Compared with $\Gamma_{A'}^{\rm eff}$, the collision rate $\Gamma_\psi^{\rm eff}$ will be much 
smaller than $0.1 {\rm Gyr}^{-1} $, due to the multiple scattering suppression. 
Therefore, DM $\psi$ should behave similarly to the collisionless DM, unless $g'$ is big enough to 
form core profiles by $\psi$-$\psi$ self-scattering. 

With the appropriate collision rate, there is another issue that whether colliding with other species
can lead to a core profile. Firstly, it has been shown in \cite{Schutz:2014nka} that the dark matter 
with an excited state can potentially solve the small structure problems. Secondly, in a microscopic
picture, many $A'$ collides with the same $\psi$. Therefore, the $\psi$ serves as a bridge, which 
effectively mediates the kinetic energy exchange between different $A'$.
Since the momentum of $\psi$ itself does not change significantly because of the random walk suppression, and
moreover it has a small fraction in relic abundance, its total momentum and kinetic energy are
subdominant comparing with the light dark matter. As a result, $\psi $ itself is not important
in the structure formation and the net effect is the heat change between different $A'$.
Finally, there is another semi-analytical way to understand this by using the Boltzmann equations.
Following \cite{Ahn:2004xt, Walker:2009zp, Oh:2010ea}, the core profile of dark matter can be determined by requiring hydrostatic equilibrium.
For the self-interacting dark matter, it is achieved by the kinetic equilibrium, and the 
core density profile is the solution with the proper boundary conditions.
For the two-component dark matter, after reaching the kinetic equilibrium, the equilibrium equation 
is the same for the dominant dark matter, so that the core density profile is the solution.
\\

2) \textit{$\psi$ dominant}, $F_{\psi} \approx 1 \gg F_{A'}$. 
The \textit{effective} collision rate for $\psi$ in Dwarf galaxies is:
\begin{align}
& \Gamma_{\psi}^{\rm eff} \approx  0.3 {\rm Gyr}^{-1}  \left(\frac{F_{A'}}{0.01} \right)^2 
\left( \frac{g'}{10^{-5}} \right)^4 \left( \frac{m_{A'}}{10^{-19}\rm eV}\right)^{-2}   \label{eq:GammaPsinum}  \\
& \times \left( \frac{m_{\psi}}{1 \rm GeV}\right)^{-5} \left( \frac{v_{\rm rel} }{10 \rm{km/s} } \right) \left( \frac{v_0}{10 \rm{km/s}} \right)^{-3}
\left(\frac{\rho_{\rm DM}}{0.1{\rm M}_{\odot}/ {\rm pc}^3}\right)^2 \nonumber .
\end{align}
This rate is $m_\psi / m_{A'}$ times smaller than $\Gamma_{A'}^{\rm eff}$, such that 
the subdominant $A'$ will change its momentum in a time scale much shorter than the galactic one. 
We expect that the collision with $A'$ will cool the DM $\psi$ to have an equal partition of kinetic energy, 
since $A'$ has much larger number density, and the $A'$ kinetic energy is much smaller than $\psi$ as it starts.
It will dissipate the energy of $\psi$ at a typical time scale of $1/\Gamma_{\psi}^{\rm eff}$.
The cooling will lead to a denser and smaller core for $\psi$. 
Therefore, it is difficult for the $\psi$ dominant case to form core profiles through interactions with the dark photons. 
\newline

\section{Constraints} 
\begin{figure}
	\centering
    \includegraphics[width=0.95 \columnwidth]{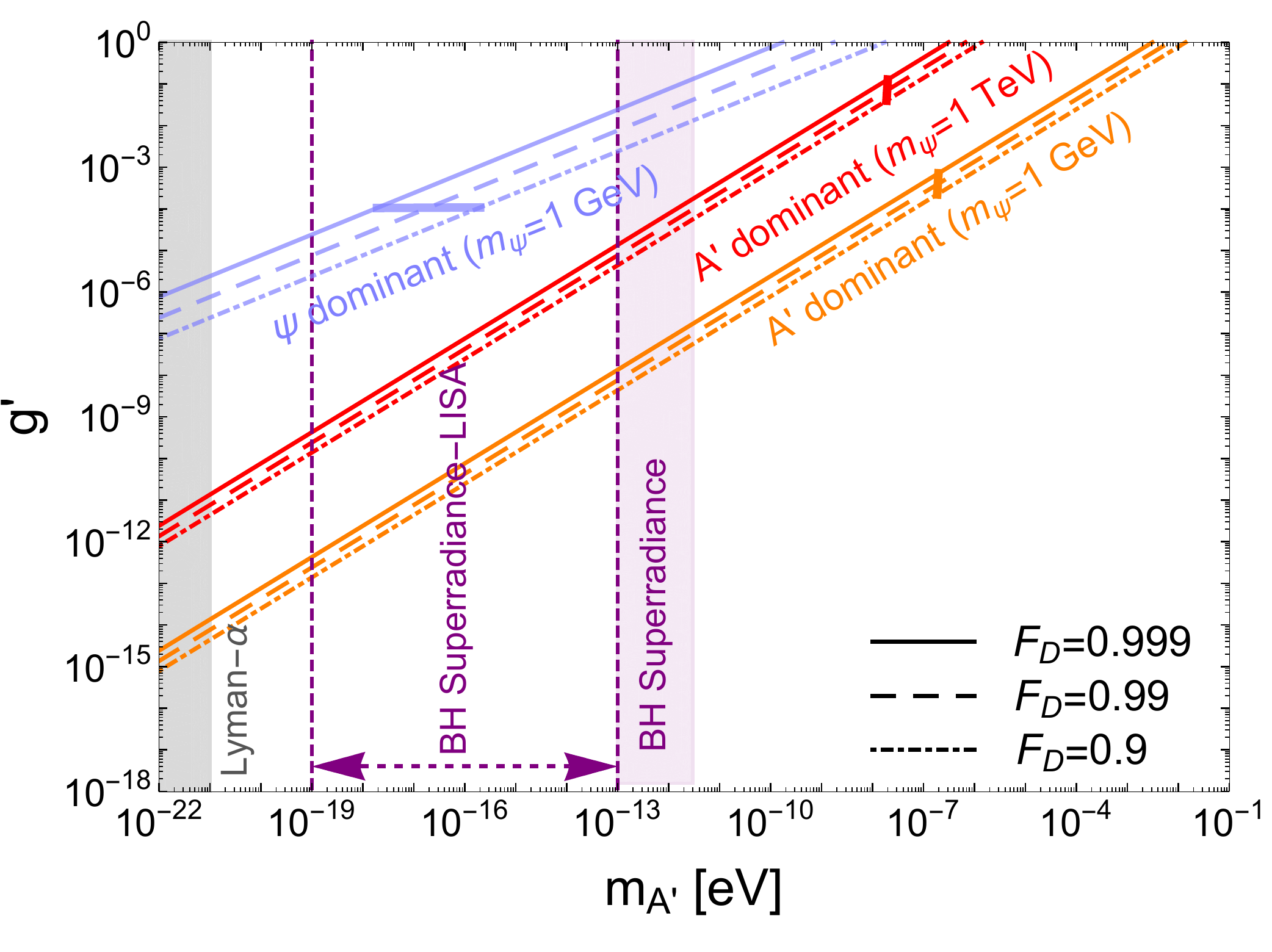} 
	\caption{
		The parameter spaces for the dominant DM component (either ultralight $A'$ or WIMP 
		$\psi$) having a core profile in typical Dwarf galaxies.
       The solid, dashed and dot-dashed lines indicate the dominant DM with density fraction $F_D = 0.9, 0.99 $ and $0.999$.
		The current black hole superradiance limits on $A'$ and the future projection from LISA  are plotted. The solid lines intersecting with the bands show the 
		required $g'$ to have $\psi$ self-scattering rate reach $0.1 {\rm Gyr}^{-1}$ in typical Dwarf galaxies.
		The Lyman-$\alpha$ constraint exclude $m_{A'} \lesssim 10^{-21}$ eV, 
		if $A'$ is dominant.
}	\label{fig:Apparameter}
\end{figure}

In Fig.~\ref{fig:Apparameter}, 
we plot the parameter spaces for the $A'$ and $\psi$ dominant cases whose
\textit{effective} interaction rate 
equals to $0.1 ~{\rm Gyr}^{-1}$ in the Dwarf galaxies. 
The ultralight vector $A'$ can be constrained by the black hole superradiance \cite{Cardoso:2018tly}, 
and the future reaches from LISA \cite{Audley:2017drz} are also plotted.
The Lyman-$\alpha$ constraint excludes $m_{A'} \lesssim 10^{-21}$ eV \cite{Menci:2017nsr, Irsic:2017yje, Armengaud:2017nkf, Kobayashi:2017jcf, Murgia:2018now, Nori:2018pka}, 
which is shaded in gray in the graph.

Furthermore, we have checked the subdominant $\psi$ self-scattering rate 
$\Gamma_\psi^{\rm self} = n_\psi \sigma_T v_{\rm rel}$, with momentum transfer
cross-section $\sigma_T$ from \cite{Tulin:2012wi} including the Sommerfeld enhancement.
The $\psi$ self-scattering rate depends on $m_\psi$, $g'$ and $F_\psi$, but not so much
on $m_{A'}$,  because $m_{A'} \ll m_{\psi} v_0$. 
We plot the corresponding $g'$ for $\Gamma_\psi^{\rm self} \sim 0.1 ~ {\rm Gyr}^{-1} $ in 
typical Dwarf galaxies as a line that intersects the band. When $g'$ is smaller than that,
$\psi$ is close to the collisionless and has a cuspy profile.
For $\sigma_T /m_{\psi}$ larger than that, it will have quite strong self-interaction, however
as long as its density fraction $F_{\psi} \lesssim 23 \%$,  it is not limited by the bullet cluster 
\cite{Randall:2007ph}.


Next, we consider the astrophysical constraints on the CoIDM model, 
including galaxy mergers~\cite{Cole:2000ex,Springel:2000qu,DeLucia:2006szx}, subhalos moving in their parent halos~\cite{Vogelsberger:2012ku} and halo shapes~\cite{Dave:2000ar, Tulin:2017ara}. 
Different from the analysis in the same halo, 
when considering the interactions between two halos,
the $ v_{\rm rel} \sim v_0 $ condition can be violated.
The final state Bose enhancement $\langle \mathcal{N}^{A'} \rangle$ depends on DM ${A'}$ velocity 
dispersion $v_0$ and the final state momentum of $A'$.
If ${A'}$ and $\psi$ come from different DM halos, with a relative velocity between the two 
halos $v_{\rm rel} \gg v_0$, the final state $A'$ will have a velocity of order $\mathcal{O}(v_{\rm rel})$
for a typical collision. 
For a Maxwellian distribution, there is an exponential suppression factor 
$e^{- v^2_{\rm rel} / v_0^2}$, such that one should substitute $ \langle \mathcal{N}^{A'} \rangle \to 
\langle \mathcal{N}^{A'} \rangle  e^{- v^2_{\rm rel} / v_0^2}$ in the collision kernels. 
When $v_{\rm rel}$ is much larger than the escape velocity of the subhalo, the Bose enhancement vanishes.
In this case, the next order results for the collision kernels $\mathcal{C}_{\psi,A'}$ needs to be considered.
These results are presented in the Appendix.

Starting with galaxy mergers, we consider the Bullet Cluster a generic 
example~\cite{Markevitch:2003at,Randall:2007ph,Bradac:2008eu,Robertson:2016xjh}.
For SIDM, the self-interacting cross-section satisfies $\sigma / m \lesssim \mathcal{O}(1)  {\rm cm}^2 / {\rm g}$, 
such that the DM halos will be separated and consistent with the gravitational
lensing observations. For CoIDM, the constraint is that the product of the momentum exchange rate and the Bullet Cluster
crossing time must be less than 1, $\Gamma^{\rm eff} t^{\rm cross} <1$. With offset $\sim 25 {\rm kpc}$ and
relative velocity $v_{\rm rel}^{\rm BC} \sim 4000 {\rm km}/{\rm s}$ \cite{Tulin:2017ara}, 
the Bullet Cluster crossing time $t_{\rm cross}$ is approximately $6\times 10^6$ years. Thus, the rate in the Bullet 
Cluster should satisfy $\Gamma^{\rm eff}_{\rm bullet} < 0.016 {\rm Gyr}^{-1}$.  
The difference between $v_{0}^{\rm BC} \sim 1000 ~ {\rm km}/{\rm s}$ \cite{Cline:2014vsa} and 
$v_{\rm rel}^{\rm BC} \sim 4000 {\rm km}/{\rm s}$ leads 
to an exponential suppression in the Bose enhancement factor  
$\langle \mathcal{N}^{A'} \rangle e^{-(v_{\rm rel}^{\rm BC}/v_{0}^{\rm BC})^2} \sim 10^{-7} \langle \mathcal{N}^{A'} \rangle$. 
Furthermore, the density of the Bullet Cluster, $\rho^{\rm BC} \sim 10^{-3} {\rm M}_{\odot} /{\rm pc}^{-3}$ 
\cite{Cline:2014vsa} is smaller than that of Dwarf galaxies. Thus, we conclude that the Bullet Cluster does not constrain the CoIDM model for the $A'$ dominant case.

When a collisions occurs between $A'$ and $\psi$ due to individual halos, the rates for $\psi$ are different from 
those in the same halo.
By neglecting the velocity dispersion of $A'$ in the other halo, $A'$ 
has monochromatic velocity $v_{\rm rel}$ in the $\psi$'s point of view. 
Therefore, the momentum change for $\psi$ in each collision can add up in the moving direction, 
since it is a head-on collision. As a result, the random walk factor $(m_\psi/m_{A'})^2$ 
is replaced by the number of head on collisions needed $\sim m_\psi/m_{A'}$. 
The collision rate is:
\begin{align}
\Gamma_{\psi}^{\rm head-on} \approx \Gamma_{\psi}^{\rm eff} \frac{m_\psi}{m_{A'}}.
\end{align}
We apply the model parameters in eq.~(\ref{eq:GammaPsinum}),
consider $v_{\rm rel}^{\rm BC}$,  $v_{\rm 0}^{\rm BC}$ in the Bullet Cluster,  
and include the exponential suppression factor $e^{-(v_{\rm rel}^{\rm BC}/v_{0}^{\rm BC})^2} $. This leads to 
$\Gamma_{\psi}^{\rm head-on} \sim 10^{13} {\rm Gyr}^{-1}$ in the $\psi$ dominant case. 
From this collision rate, it appears that the $\psi$ DM is constrained by the Bullet Cluster.
However, if the velocity distribution deviates from Maxwellian, or has a smaller escape velocity,
the final state Bose enhancement will not happen and the constraint is avoided.
We conclude that for the $\psi$ dominant case, the Bullet Cluster could be relevant. However, this is dependent 
on the velocity distribution.

The same argument in the cluster merger should be applied when subhalos travel 
in the main halos~\cite{Vogelsberger:2012ku}. 
It has been found that the main halos usually have larger velocities, and much smaller density than subhalos at 
the same position~\cite{Ghigna:1999sn, Zentner:2003yd, Diemand:2004kx, Diemand:2007qr, Springel:2008cc, Rocha:2012jg}. 
Moreover, the substructures tend to appear in the outer regions of the main halos, 
which enlarges the density difference between them~\cite{Springel:2008cc}. 
Furthermore, the Bose enhancement is suppressed by 
$e^{-(v_{\rm rel}^{\rm main}/v_0^{\rm sub})^2}$ due to the large $v_{\rm rel}^{\rm main}$. 
Each of this effects leads to the conclusion that 
the subhalo DM is not destroyed by the main halo in the $A'$ dominant case.
For the $\psi$ head-on collisions, this depends on the velocity distributions of the final states.

The last constraints that we consider are the halo shapes from the observations of the elliptical galaxies and clusters \cite{Rocha:2012jg, Tulin:2017ara}. 
It is known that the thermalization of SIDM in the center makes the halos
more spherical, while the collisionless DM has a minor-to-major axis ratio of $0.6-0.7$
\cite{Dave:2000ar}. 
In $A'$ dominant CoIDM, the subdominant $\psi$ is collisionless in the galaxy time scale considering a small $g'$, e.g. $g'$, lower than the line that intersects the band in Fig.~\ref{fig:Apparameter}. 
It is expected to have density profiles and minor-to-major axis ratios
similar to the cold DM. The scattering rate 
of $A'$ is proportional to the density of $\psi$, thus the shape tends to follow the density profile of $\psi$.
Moreover, since we fix the collision rate for $A'$ at $0.1 ~{\rm Gyr}^{-1}$ in Dwarf galaxies, the rates in
the galaxy and clusters are much smaller. Therefore, the halo shape constraints do not apply to $A'$ 
dominant CoIDM.

In summary, different from collision inside the same (sub)halo, the momentum exchange for every 
collision of $A'$ and $\psi$ between two DM halos has preferred directions due to the high velocity. 
This suppresses the final state Bose enhancement and weakens the multiple scattering requirement
for $\psi$. As a result, the $A'$-dominant case is safe from these constraints, while $\psi$
has inconsistent with them. However, this is  dependent on the velocity distribution. 
\newline

\section{Discussions}
The following are several points pertinent to CoIDM.

1) The self-interaction of dark photons $A'A' \to A'A'$ has a rate much smaller than the 
interaction rate with $\psi$. Note the reaction and back-reaction cancels the leading term of $\mathcal{N}_{A'}^4$,
its rate can be estimated as $n_{A'} \langle \sigma v \rangle_{A'A'\to A'A'} \langle \mathcal{N}^{A'}  \rangle$ that
it is only enhanced by $\langle \mathcal{N}^{A'} \rangle$ linearly. Given that the self-interaction cross section is about ${\alpha'}^{4} m_{A'}^6/m_\psi^8$, 
with $\alpha' = {g'}^2/(4 \pi)$, one can check explicitly the rate is much smaller than $0.1 {\rm Gyr}^{-1}$, 
due to the great suppression by the large $\psi$ mass.

2) Similar to the axion in the non-relativistic limit, there should be soliton solutions for vector DM.
Using typical central density for Dwarf galaxies, one 
can determine the soliton mass as
$\sim 2\times 10^6 M_{\odot} \left(10^{-19} {\rm eV}/m_{A'} \right)^{3/2}$ \cite{Hui:2016ltb}, 
much smaller than Dwarf galaxy mass. Therefore, 
the soliton solutions do not affect the core profiles given by the CoIDM model in the 
galaxies. 

3) The kinetic theory, eq.~(\ref{eq:boltz}), is valid if the rate for $\psi A' A' \to \psi A' A'$   
is subleading. In the parameter spaces we have analyzed, this was easy to satisfy. Moreover, the $2 \to 3$ or $3 \to 2$ processes are 
kinetically forbidden or do not have the final state enhancement. 

4) The wavelength of $A'$ is much larger than the separation of $\chi$ particles. One should consider
the coherent scattering of many $\chi$. We assume DM is charge symmetric. If the total number of
DM is $N_{\psi}$, one should have a net charge of about $\sqrt{N_\psi}$ and the rate should be proportional
to $N_{\psi}$. Therefore, the rate would be same as in particle scattering, shown in eq.~(\ref{eq:Apsics}). 

5) In the early universe, the collision rate between $A'$ and $\psi$
is much larger, due to the higher number densities and lower velocities. 
However, the interaction will only redistribute the energy in the dark sector
when $A'$ and $\psi$ DM are non-relativistic and decoupled from the plasma. 

6) The dark photons will have the plasma mass $\sim g' \sqrt{\rho_{\psi}/m_{\psi}^2}$, 
but this is smaller than the bare $A'$ mass in the parameter space in Fig.~\ref{fig:Apparameter}. 

7) We have considered the Thompson scattering between $A'$ and $\psi$, and
all the exotic features coming from the existence of high number density for $A'$ in the space. 
Such collisions will lead to the same kinetic energy of $A'$ and $\psi$, thus cool $\psi$ due to the large
number of $A'$. However, there could also be absorption of $A'$ which could heat the plasma \cite{Dubovsky:2015cca}, 
where the rate of absorption is roughly the collision rate of $\psi$ themselves.
In Fig.~\ref{fig:Apparameter}, in the region below the solid intersection lines, the friction induced by
plasma collision is not significant enough to change $A'$ energy density.
Moreover, at large $g'$, $\psi$ can dissipate its energy via $A'$ emission 
$\psi \psi \to \psi \psi A'$~\cite{Essig:2018pzq, Chang:2018bgx}, 
and we found that the effect is small at both early universe and late time for 
the parameter space we considered. 
In summary, the $g'$ is small enough that both heating via $A'$ absorption or cooling
via $A'$ emission are negligible. 
Moreover, if we change our model so that ultralight DM is a scalar field $\phi$, with interaction with
$\psi$ via a higher dimensional operator $|\phi^2| \bar{\psi} \psi/\Lambda$, both effects 
can be avoided because the $\phi$ number is conserved.

8) In addition to the ultralight dark photons as DM, one could choose both DM $\psi_{1,2}$ as
WIMP with mass hierarchy $m_{1} \gg m_2$, (for small mass difference, see \cite{Schutz:2014nka}). 
When the self-interaction can be neglected, 
the subdominant DM can help the dominant one to thermalize. Although there is no occupation 
number enhancement, one still needs to consider the random walking suppression for the heavier DM 
particles and the forward-backward scattering cancellation, which leads to significant differences 
from SIDM. 
We list the relevant rates for this scenario in the Appendix. 
\newline

\section{Conclusion}
We introduce a subdominant DM $\psi$ to assist the thermalization of ultralight dark photons, 
which could help the latter to form a core profile, even with a dark photon mass larger than $10^{-21}$ eV. 
Therefore, the usual constraint from Lyman-$\alpha$ does not apply and the small structure issues
for galaxies are solved. Furthermore, the Bullet Cluster bound is evaded easily due to the reduced Bose enhancement factor.
The scattering rates for $\psi$ and $A'$ are enhanced by the large occupation number of $A'$,  
and suppressed by the forward-backward scattering cancellation. Meanwhile, the rate for $\psi$ is suppressed by the multiple scattering requirement. 
Considering the limits from the merger of galaxies, the $\psi$ dominant case is possibly
constrained, but this depends on the velocity distributions, while in $A'$ dominant case, they are safe 
from these constraints. N-body simulations would be helpful for making more concrete statements on the small scale issues in galaxies.
\newline

\begin{acknowledgements} \noindent {\bf \textit{Acknowledgment.}}
	We thank Yang Bai, Xiaoyong Chu, Dan Hooper, Yonatan Kahn, Doojin Kim, Joachim Kopp, Andrew Long, Ian Low, Matthew Low, Dam Son, Lian-Tao Wang, Neal Weiner, Bin Wu, Yi Yin, Haibo Yu and Yue Zhao for useful discussions.  
	JL is supported by the Oehme Fellowship.
	XPW is supported  by the U.S. Department of Energy under Contract No. DE-AC02-06CH11357.
	WX is supported by the European Research Council grant NEO-NAT.
\end{acknowledgements}


\section*{Appendix}
To be complete, we show here the next-leading term in the collisional kernel of eq. 3, 
when no $\mathcal{N} \gg 1$ is assumed. 
Without the large occupation number enhancement, it becomes the leading term in the case that both DM components are particles. It is useful for the two-component DM model 
which has mass hierarchy, for example $\chi_{1,2}$ as Dirac DM with $m_1 \ll m_2$.
We only allow interaction between them and neglect their self-interactions. 
Interestingly, the forward backward cancellation only appears in heavier DM 
$\chi_2$ scattering rate. To be detailed, the collision kernel is now proportional to 
$ \mathcal{N}^{\chi_1}_{p_1} \mathcal{N}^{\chi_2}_{k_1} - \mathcal{N}^{\chi_1}_{p_2} \mathcal{N}^{\chi_2}_{k_2}$, 
and we can separate it into 
\begin{align}
\mathcal{N}^{\chi_1}_{p_1} \mathcal{N}^{\chi_2}_{k_1} - \mathcal{N}^{\chi_1}_{p_2} \mathcal{N}^{\chi_2}_{k_2} 
\approx
\mathcal{N}^{\chi_2}_{k_1} \left( \mathcal{N}^{\chi_1}_{p_1} - \mathcal{N}^{\chi_1}_{p_2} \right) +
\frac{m_{\chi_1}}{m_{\chi_2}} \mathcal{N}^{\chi_2}_{k_1} \mathcal{N}^{\chi_1}_{p_2} \nonumber
\end{align}
For the collision rate of ${\chi_1}$, it integrates over $\int d\vec{p}_2 \int d\vec{k}_1 \int d\vec{k}_2 $ and it can be 
shown that
$\int d\vec{p}_2 \mathcal{N}^{\chi_1}_{p_1} - \int d\vec{p}_2 \mathcal{N}^{\chi_1}_{p_2} \approx \mathcal{O}(1) \mathcal{N}^{\chi_1}$.
However, for the collision rate of ${\chi_2}$,  it integrates over $\int d\vec{p}_1 \int d\vec{p}_2 \int d\vec{k}_2 $ and 
the term
$\int d\vec{p}_2 d\vec{p}_1 \mathcal{N}^{\chi_1}_{p_1} - \mathcal{N}^{\chi_1}_{p_2} $ vanishes in the leading term. Therefore, ${\chi_2}$ receives $\frac{m_{\chi_1}}{m_{\chi_2}}$ suppression. 
Thus, we list the collision rate for ${\chi_1}$ and ${\chi_2}$,
\begin{align}
\Gamma_{\chi_2}^{\rm eff} & =  \frac{ m_{\chi_1}^2} {m_{\chi_2}^2} \frac{1}{\mathcal{N}^{\chi_2}_{\bf{k_1}}}\partial_t \mathcal{N}^{\chi_2}_{\bf{k_1}} \simeq n_{\chi_1} \left\langle \sigma v \right\rangle_{{\chi_1}{\chi_2} }    \frac{m_{\chi_1}^3}{m_{\chi_2}^3} , \nonumber
\\
\Gamma_{\chi_1}^{\rm eff} & = \frac{1}{\mathcal{N}^{\chi_1}_{\bf{p_1}}} \partial_t \mathcal{N}^{\chi_1}_{\bf{p_1}} \simeq n_{\chi_2} \left\langle \sigma v 
\right\rangle_{{\chi_1}{\chi_2}}   . \nonumber
\end{align}

\bibliography{ref}

\bibliographystyle{JHEP}

\end{document}